\documentclass{evn2004}
\usepackage{txfonts}
\begin{document}
\setcounter{page}{221}

   \title{Probing the polarization characteristics of SS433 on mas scales} 

   \author{Z. Paragi\inst{1}, R.C. Vermeulen\inst{2}, D.C. Homan\inst{3}, J.F.C. Wardle\inst{4},
           I. Fejes\inst{5}, R.T. Schilizzi\inst{6,7}, R.E. Spencer\inst{8}, A.M. Stirling\inst{8} 
          }

   \institute{ \inst{1}JIVE, Postbus 2, 7990\,AA Dwingeloo, The Netherlands,
               \inst{2}ASTRON, Postbus 2, 7990\,AA Dwingeloo, The Netherlands,
	       \inst{3}Denison Univ., Physics Dep., Granville, OH 43023, United States,
               \inst{4}Brandeis Univ., Physics Dep., MS 057, P.O. Box 549110, Waltham, MA 02454-9110, United States, 
               \inst{5}F\"OMI SGO, P.O. Box 585, H-1592 Budapest, Hungary,
               \inst{6}International SKA project Office, 7990\,AA Dwingeloo, The Netherlands,
               \inst{7}Leiden Obs., P.O. Box 9513, 2300 RA Leiden, The Netherlands,
	       \inst{8}JBO, Jodrell Bank, Macclesfield, Cheshire, SK11 9DL, United Kingdom
             }

   \abstract{

We present a status report of dedicated circular polarization (CP) global VLBI observations
of SS433. The total intensity image of the source is presented. We do not detect
linear polarization in the core region of the source. We estimate the errors in D-terms
for the EVN, and its effect on the final right-left circular polarization gain calibration. 
The data from this and earlier observations are analyzed with the "zero-V" self-calibration 
technique, in order to obtain indications for detectable CP in different regions of the source. 
Our data show no evidence for strong gyro-synchrotron (GS) emission on the size scale probed
by our beam.

 
   }

   \authorrunning{Paragi et al.}
   \titlerunning{SS433 polarization on mas scales}
   \maketitle
%

\section{Introduction}

SS433, is a microquasar (radio-jet X-ray binary system).
It ejects antiparallel beams at a near relativistic velocity of 0.26c
(see Vermeulen \cite{RCV95} and references therein). 
High resolution EVN experiments showed that the radio core of the source 
has double core-wing morphology (Vermeulen et al. \cite{RCV93}).
Recent observations with the VLBA and the
EVN revealed that the compact jet region is similar in many respects to the
radio cores observed in quasars. These observations also showed
evidence for an \lq\lq equatorial outflow" in the system, unprecedented in
radio-jet sources before (Paragi et al. \cite{ZP99a}, \cite{ZP02}). 
The equatorial emission region was also imaged in an experiment 
using the phased VLA, MERLIN and the VLBA (Blundell et al. \cite{KB01}).

The beams of SS433 are unique in the sense that they contain baryonic matter
evidenced from Doppler-shifted spectral lines observed in the optical
regime (e.g. Margon \cite{BM84} and references therein) and X-rays (e.g. Kotani et al. 
\cite{KOT96}). They also contain highly relativistic electrons (with Lorentz factors
$\gamma\sim 300$) radiating by the synchrotron process. The low-energy
cutoff of this electron population -- just as in other microquasars
or quasars -- is not known. But the contribution of 1--100 MeV electrons
($\gamma=2-200$) to the total luminosity may give rise to detectable amount
of circular polarization (CP) in the source (Spencer \& McCormick \cite{RES03}).

Recently it was demonstrated -- in spite of the difficulties with calibration --
that circular polarization can be detected in quasars with the VLBI technique, using the
VLBA array (Wardle et al. \cite{WAR98}). The CP emission is related to the radio core
of the sources that is assumed to be a transition region where radiation becomes
optically thin. In microquasars CP emission was detected in GRS1915+105
(Fender et al. \cite{RPF02}) and SS433 (Fender et al. \cite{RPF00}) with 
connected-element interferometers. There is no high resolution observation 
obtained to date that shows circular polarization in these sources.



\section{The global VLBI observations}

We carried out global VLBI observations of SS433 on 29 May 2003, including several 
calibrator
sources dedicated for CP calibration. The array consisted of the western EVN, 
the VLBA, the Green Bank Telescope, and a single dish of the VLA. The observing
frequency was 1.6~GHz. The recording rate was 256 Mbps using 2 bit sampling.
There were two 16~MHz channels recorded in both RCP and LCP polarizations.
This setup was used because we wanted to maximize the channel sensitivity,
which simplifies the fringe-finding procedure a great deal, given the target source
can be detected in data from a single channel. All channels were related to the
lower sideband of the video converters. This, coupled with the fact that the experiment
had a fan-out 1:4 caused complications in correlation. The correlator control software at JIVE
had to be upgraded in order to handle this mode properly.


The data calibration was carried out in AIPS. In a first attempt, we used the
amplitude calibration information provided by the EVN Pipeline (Reynolds, Garrett \& Paragi \cite{Pipe}). 
However we realized
that much of the $T_{\mathrm sys}$ data were missing for the Green Bank Telescope.
To avoid unnecessary flagging of data due to missing calibration information, we inserted
$T_{\mathrm sys}$ values in each scan, typical for the sources. In the case of Noto and 
Torun we found that 
the stokes $V$ values were in excess of 10$\%$. The problem with calibrating these stations is
under investigation.
These three stations could not be included in our CP analysis.

For correcting the polarization leakage D-terms and calibrating the gain of the $R$ and $L$ systems, 
we included five bright sources ($\sim$ 1~Jy) in the experiment: OQ208, 3C345, NRAO512,
J1832+1357, J2002+4725. Circular polarization has not been detected in these sources to date. 
After parallactic angle correction we run task LPCAL in AIPS. This was done for all
calibrators, to estimate how accurate our D-term calibration was (Table \ref{Dterms}). 
Eventually we used the D-terms from OQ208, which is practically unpolarized.

After applying the D-terms we imaged all the calibrator sources in Difmap. The next step will be the proper
calibration of the right and left handed system gains, using all the calibrator sources.
The idea is that, on average, these have no circular polarization. Any source that has significant 
CP with respect to the others has to be left out from this procedure (Homan \& Wardle \cite{H&W99}). 
This work is still under progress.


\begin{table*}
\caption[]{Amplitudes ($A_{\rm R}$, $A_{\rm L}$) and phases ($\phi_{\rm R}$, $\phi_{\rm L}$) of the polarization leakage D-terms for the right 
and left systems, respectively. Amplitudes are given in percentages. The 1~$\sigma$ errors were estimated from D-term values determined independently 
for each calibrator sources. For the US array, only the three brighter calibrators (3C345, OQ208, NRAO512) were used. }
\label{Dterms}
\begin{tabular}{lrrrrlrrrr}
\hline
\noalign{\smallskip}
  Code & $A_{\rm R}\pm\sigma_{\rm R}$ &  $\phi_{\rm R}\pm\sigma_{\rm R}$ & $A_{\rm L}\pm\sigma_{\rm L}$ &  $\phi_{\rm L}\pm\sigma_{\rm L}$ 
& Code & $A_{\rm R}\pm\sigma_{\rm R}$ &  $\phi_{\rm R}\pm\sigma_{\rm R}$ & $A_{\rm L}\pm\sigma_{\rm L}$ &  $\phi_{\rm L}\pm\sigma_{\rm L}$  \\ 
\noalign{\smallskip}
\hline
\noalign{\smallskip}

  Ef  & 3.76$\pm$0.39  & -80.33$\pm$8.06   & 7.53$\pm$0.67 & -116.24$\pm$  6.55 
& SC  & 2.70$\pm$0.62  & -51.65$\pm$199.86 & 1.51$\pm$0.25 &  -55.82$\pm$  3.00  \\
 Mc   &11.00$\pm$0.62  &   7.20$\pm$3.11   &17.95$\pm$0.68 &  105.08$\pm$156.55 
& HN  & 2.89$\pm$0.54  & -44.43$\pm$12.17  & 1.45$\pm$0.48 &  142.82$\pm$ 14.24  \\
 Wb   & 5.94$\pm$1.21  &  34.53$\pm$193.98 & 4.84$\pm$0.65 & -112.91$\pm$  4.23 
& NL  & 2.77$\pm$0.67  &  64.38$\pm$191.72 & 2.34$\pm$0.69 &   49.87$\pm$ 13.92  \\
 On   & 4.36$\pm$0.57  & -49.17$\pm$7.83   & 3.72$\pm$0.50 &   -3.76$\pm$  2.83 
& FD  & 1.60$\pm$0.39  & 160.52$\pm$ 23.53 & 3.33$\pm$0.11 &  -10.27$\pm$  8.68  \\
 Tr   & 7.76$\pm$0.37  & 117.31$\pm$5.49   & 6.87$\pm$0.38 &   69.32$\pm$  3.86 
& LA  & 1.73$\pm$0.98  & -73.89$\pm$7.88   & 2.39$\pm$0.28 &   51.31$\pm$199.98  \\
 Nt   & 2.45$\pm$0.36  & -47.32$\pm$7.88   & 3.07$\pm$0.22 &   59.41$\pm$ 12.45 
& PT  & 3.70$\pm$0.16  &-152.38$\pm$16.23  & 2.03$\pm$0.21 &  161.64$\pm$ 14.30  \\
 Hh   & 5.57$\pm$1.15  & 139.52$\pm$14.84  & 7.44$\pm$0.43 &   42.84$\pm$  7.87 
& KP  & 2.98$\pm$0.60  &-112.45$\pm$10.74  & 0.78$\pm$0.53 &   77.82$\pm$ 16.34  \\
 Gb   & 3.19$\pm$0.71  &  -8.91$\pm$23.84  & 4.15$\pm$1.49 & -151.17$\pm$ 25.18 
& OV  & 1.72$\pm$0.98  & 139.58$\pm$20.44  & 0.94$\pm$0.66 &   52.49$\pm$ 25.55  \\
 Y1   & 2.59$\pm$0.69  & -52.45$\pm$12.81  & 2.51$\pm$0.51 & -155.05$\pm$ 12.07 
& BR  & 2.79$\pm$0.58  &-150.93$\pm$22.04  & 3.69$\pm$0.73 &   56.54$\pm$  7.72  \\
      &                &                   &               &                   
& MK  & 2.20$\pm$0.20  & -63.11$\pm$ 51.42 & 1.89$\pm$0.96 &   17.92$\pm$ 16.40  \\

\noalign{\smallskip}
\hline
\end{tabular}
\end{table*}

   \begin{figure*}
   \centering
   \vspace{450pt}
   \includegraphics{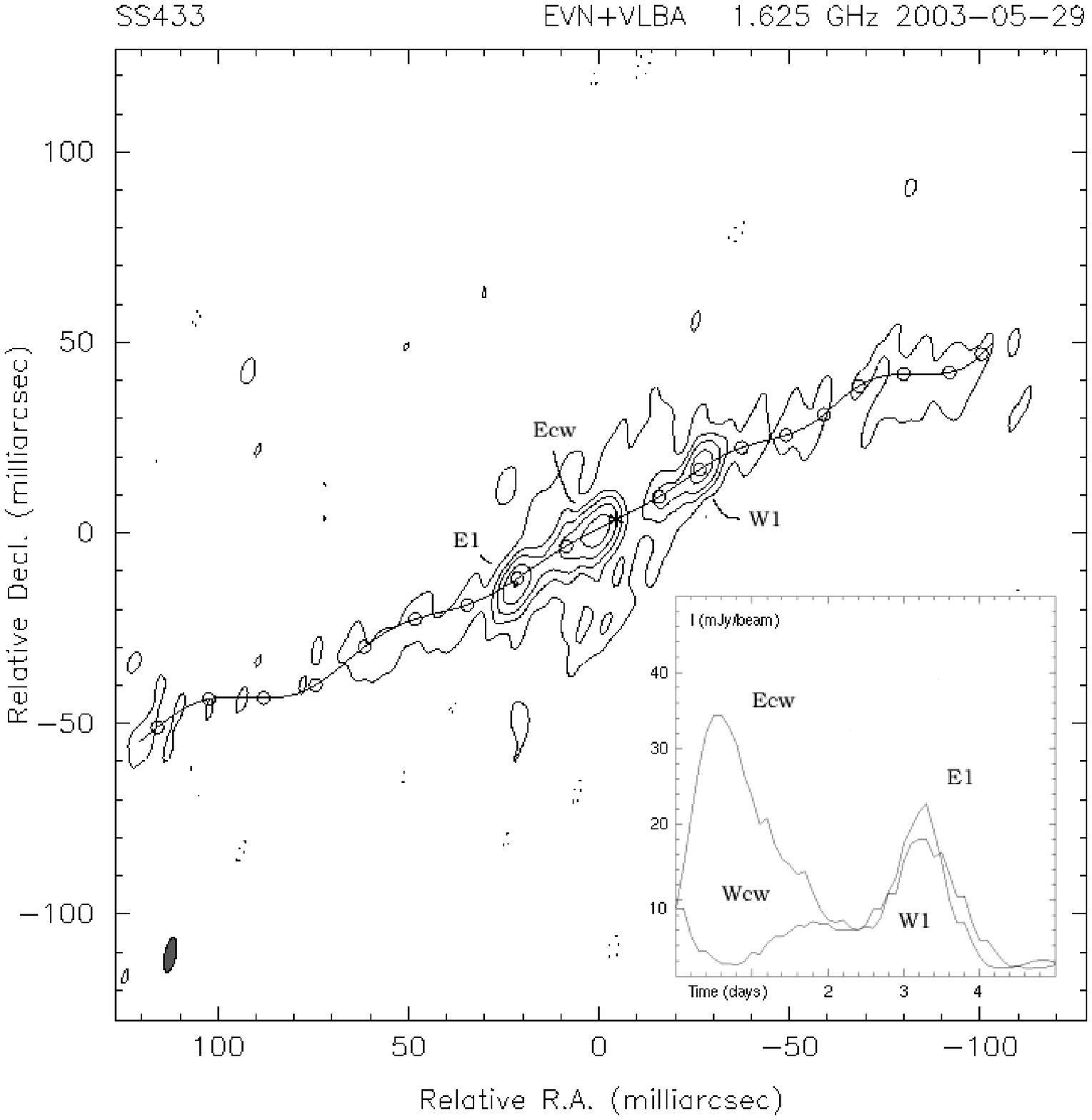}
   \caption{Uniform weighted global VLBI map of SS433 on 29 May 2003, at 1.6 GHz.
            The contour levels are -2, 2, 4, 8, 16, 32, 64\% of the peak brightness 
            of 35.3 mJy/beam. The restoring beamsize is 9.19$\times$2.98 mas, with the
            major axis oriented at PA=$-11.3$. Overlaid is the model of the precessing 
            beams. The star indicates the kinematic centre, the open circles indicate
            the plasmon ejection time at every 1.6 days. The inset shows the 
            brightness profiles of the approaching and the receding jets, with the 
            ejection time -- according to the precessional model -- indicated.
            \label{fig:SS433img}
           }
    \end{figure*}


\section{Polarization purity of the array and possible effects on the CP calibration}  

%

Determining circular polarization with VLBI is the art of calibrating the gains for
the left and right handed systems properly (Homan \& Wardle \cite{H&W99}). For the
VLBA, the effect of polarization leakage is rather small. However for the EVN it
is worth estimating how the D-terms might affect the CP calibration in an experiment.
 
The polarization-leakage D-terms of the telescopes are listed in Table~\ref{Dterms}.
The statistical errors were estimated from solutions using different calibrator sources.
These errors are in fact upper limits, because there is an uncertainty in the polarization
structure of the calibrators as well. We used the solutions from OQ208, which is not
affected by this error since it is unpolarized. The errors were in general below the $1\%$ level 
in D-term amplitude for the array. Green Bank and Westerbork had slightly
higher errors. These stations may have had a variable D-term throughout the experiment.
The scatter in D-term phases was order of $10\%$ for the EVN, and order of
$20\%$ for the US array. There are a couple of stations where very large phase errors
are listed. In these cases one or two phase solutions had the wrong sign, while their
absolute value was close to the other ones. It is not clear what causes this uncertainty
of phase solutions in LPCAL. 

The circular polarization (stokes $V$) is formed from the ideal $RR$ and $LL$ correlations 
as $V=(RR-LL)/2$. 
The measured quantities are described by the leakage term model (Cotton, \cite{Cot93}). 
For baseline $mn$, $r_{mn}^{RR}$ and $r_{mn}^{LL}$ can be written as: 


\begin{eqnarray}
\label{eqn:ellipticity}
r_{mn}^{RR} & = & g_{m}^{R} g_{n}^{R*} [ e^{-j(\alpha_{m}-\alpha_{n})} \cdot RR 
                  + D_{m}^{R}  e^{j(\alpha_{m}+\alpha_{n})} \cdot LR   \nonumber \\ 
            &   & \hspace{10mm}  \mbox{} + D_{n}^{R*} e^{-j(\alpha_{m}+\alpha_{n})} \cdot RL 
                  + D_{m}^{R} D_{n}^{R*} e^{j(\alpha_{m}-\alpha_{n})} \cdot LL ]   \\
r_{mn}^{LL} & = & g_{m}^{L} g_{n}^{L*} [ e^{j(\alpha_{m}-\alpha_{n})} \cdot LL 
                  + D_{m}^{L}  e^{-j(\alpha_{m}+\alpha_{n})} \cdot RL   \nonumber \\ 
            &   & \hspace{10mm}  \mbox{} + D_{n}^{L*} e^{j(\alpha_{m}+\alpha_{n})} \cdot LR 
                  + D_{m}^{L} D_{n}^{L*} e^{-j(\alpha_{m}-\alpha_{n})} \cdot RR ], 
\end{eqnarray}

where $\alpha_{m}$ is the parallactic angle at station $m$, $g$ is the gain at the specified polarization,
and * denotes complex conjugate. The measured stokes $V$ becomes

\begin{eqnarray}
\label{eqn:stokesV}
r_{mn}^{V} & = & [ e^{-j(\alpha_{m}-\alpha_{n})} (g_{m}^{R} g_{n}^{R*}-g_{m}^{L} g_{n}^{L*} D_{m}^{L} D_{n}^{L*}) \cdot RR    \nonumber \\
 &  & \mbox{} +  e^{j(\alpha_{m}+\alpha_{n})}  (g_{m}^{R} g_{n}^{R*} D_{m}^{R}-g_{m}^{L} g_{n}^{L*} D_{n}^{L*})  \cdot LR    \nonumber \\
 &  & \mbox{} +  e^{-j(\alpha_{m}+\alpha_{n})} (g_{m}^{R} g_{n}^{R*} D_{n}^{R*}-g_{m}^{L} g_{n}^{L*} D_{m}^{L})  \cdot RL    \nonumber \\
 &  & \mbox{} +  e^{j(\alpha_{m}-\alpha_{n})}  (g_{m}^{R} g_{n}^{R*} D_{m}^{R} D_{n}^{R*} - g_{m}^{L} g_{n}^{L*})\cdot LL ]/2.    
\end{eqnarray}

Equations (\ref{eqn:ellipticity}-\ref{eqn:stokesV}) show the complex way the D-terms are 
related to the station gains.  For the simplest case of a non-polarized source 
($RL=LR=0$, $RR=LL=I$), the uncorrected D-terms would change the effective gain on a 
baseline in e.g. $RR$ polarization in the following way: 
$ G_{mn}^{R} = g_{m}^{R} g_{n}^{R*} ( e^{-j(\alpha_{m}-\alpha_{n})}+ e^{j(\alpha_{m}-\alpha_{n})}D_{m}^{R} D_{n}^{R*})$. 
The errors in stokes $V$ would be in the order of $D_{m} D_{n}$. In our case, the D-terms are 
calibrated to better than $1\%$, and so the effect of any D-term errors is at the level of 
$\sim 0.01\%$. Note that the errors may be larger for strongly polarized sources. 
In this case, we cannot neglect the $D\cdot RL$ and $D\cdot LR$ terms. Also note that a station
with a variable $D_{m}$ at the $\sim1\%$ level introduces variations in the observed CP at 
the $1\%\times D_{n}$ level on baseline $mn$. In the case of the EVN, this means stokes $V$ variations
at levels of $0.1\%$, which is not negligible compared to the CP observed in the sources.

\section{The core-region of SS433}


The stokes $I$ map of SS433 is shown on Fig.~\ref{fig:SS433img}. We applied uniform weighting
to the data to make the best possible use of the global array in terms of resolution.
Overlaid on the map can be seen the precessional model of the source. We tried to find
the kinematic centre for the moving jet components E$_{1}$ and W$_{1}$, assuming they were
ejected at the same time, and varied the precession phase to get the best visual 
fit to the jet structure on the map. 
The observed precession phase was $\Psi=0.72$ (using the model from Vermeulen \cite{RCV89}). 
At this precession phase the eastern jet
is the approaching, and the western jet is the receding one. 
The kinematic centre, where the binary stellar system is located, was found to be off the
brightest component. Earlier results showed (Paragi et al. \cite{ZP99a}) that the kinematic  
centre is located mid-way between the approaching and the receding core-jets. That region
of the source is known to be heavily free-free absorbed by a spherically non-symmetric
medium (Stirling, Spencer \& Watson \cite{SSW97}; Paragi et al. \cite{ZP99a}). Our current
result seems to be a bit different. At this epoch the brightness asymmetry between the
approaching and the receding core-jet is larger than usual, and the E$_{\rm cw}$ feature 
seems to be closer to the kinematic centre than in previous observations.

There is emission surrounding the core region of the jets. It is more evident on the 
naturally weighted image (not shown). This feature is well known from previous
observations (Blundell et al. \cite{KB01}, Paragi et al. \cite{ZP99a}). One possible explanation
is that this equatorial emission region is related to a non-relativistic outflow from 
the system (Paragi et al. \cite{ZP02}). The brighter equatorial features at $PA\sim 174$ and 
$-25$ degrees may indicate enhancements in the equatorial flow, which may have resulted
in an increase in the free-free optical depth in the line of sight to the receding core-jet
component.

We detect no linear polarization in SS433 on milliarcsecond scales. This result is consistent
with earlier VLBI observations (Paragi et al. \cite{ZP99b}). The upper limit to the fractional
polarization in the E$_{\rm cw}$ feature (estimated by dividing the 3 $\sigma$ noise on the $Q$ and 
$U$ maps with the peak brightness in stokes $I$) is about $0.5\%$. The SS433 beams are known to be 
polarized on larger scales (e.g. Hjellming \& Johnston \cite{HJ81}).
Depolarization of the jets
may occur in the same medium that is responsible for free-free absorption in the core.
However the beams are depolarized up to scales of 100~mas, whereas free-free absorption
occurs only in the innermost parts of the jet.  


\section{Circular polarization in SS433}

Circular polarization in SS433 was discovered by Fender et al. (\cite{RPF00}) with the
Australian Compact Array (ATCA). They observed a steep spectral index for the emission with
$V\propto \nu^{-0.9\pm0.1}$. However, the fractional polarization could not be determined
because the CP emission could not be resolved by ATCA. McCormick et al. (\cite{GMcC03})
monitored the source with ATCA for several months and found a long-term sign reversal
in CP. The two main processes that may explain the observations is the synchrotron mechanism
itself, and linear-to-circular polarization conversion (Kennett \& Melrose \cite{LPCP}). 
Both of these processes require the presence of low-energy electrons. The predicted spectrum
of the fractional circular polarization is $m_{c}\propto \nu^{-1/2}$ in the former case,
while the spectral index maybe between $-1$ and $-3$ in the conversion models (see Fender 
et al. \cite{RPF00} and references therein). Thus, if CP is detected on mas scales in SS433,
multi-frequency high resolution observations may help us to find out which process is in 
action.

The location of the emitting region is also unclear at present. It can be the base of the
jet as in quasars (Homan, Attridge \& Wardle \cite{HA&W01}), the larger scale radio beams, 
or the extended equatorial emission region. Spencer and McCormick (\cite{RES03}) suggested that the 
gyro-synchrotron (GS) process maybe effective as well if there is a significant fraction of 
electrons with Lorentz factor $\gamma\sim 30$. This could result in very high fractional
circular polarizations. A good candidate region for this process to work is the equatorial
emission for which thermal (Blundell et al. \cite{KB01}) and synchrotron (Paragi et al.
\cite{ZP99a}, \cite{ZP02}) origins have been proposed so far, none of which was fully 
confirmed.   

An easy check for the GS scenario is self-calibrating the $R$ and $L$ systems with the total
intensity model of the radio beams (i.e. assuming no circular polarization in the source). 
One may expect that some CP emission would show up 
at the location of the equatorial emission if the fractional CP was indeed high. We performed
this test with our present data and global VLBI data taken in 1998, when the peak brightness
of the equatorial region was the highest observed to date, around 8~mJy/beam. We see no 
indication of very high fractional CP from the equatorial emission,
which means that we have no evidence for strong GS emission on the size scale probed by our beam.

\section{Conclusions}

Our observations showed that the receding core-jet of the system was fainter than usual. 
This may be due to a larger free-free optical depth in the equatorial outflow at
the epoch of our observations. 
There is no linear polarization detected in SS433 on mas scales, with an upper 
limit to the fractional polarization of $0.5\%$ in the core. The polarization leakage
D-terms in our experiment were shown to be corrected to levels where they do not
affect CP calibration significantly. Our early conclusion about the CP emission detected
by others is that it is probably not due to the GS process. The nature of the equatorial
emission remains unclear.

\begin{acknowledgements} 

We would like to thank Chris Phillips to include our special observing mode in the 
correlator control software at JIVE. We are very grateful to Bob Campbell, who 
not only helped the PI during correlation, but also worked very hard on fixing the problems with 
the experiment. ZP acknowledges partial financial support from the Hungarian 
Scientific Research Fund (OTKA, grant No. T 046097). 
The National Radio Astronomy Observatory is operated by 
Associated Universities, Inc. under a Cooperative Agreement with the 
National Science Foundation.
The European VLBI Network is a joint facility of European, Chinese, 
South African and other radio astronomy institutes funded by their 
national research councils.
This research was supported by the European Commission's I3 Programme
RadioNet, under contract No.\ 505818.
\end{acknowledgements}

\end{document}